# Cell-format-dependent mechanical damage in silicon anodes

Marco-Tulio F. Rodrigues,[1*] Sathish Rajendran,[1] Stephen E. Trask,[1] Alison R. Dunlop,[1] Avtar Singh,[2] Jeffery M. Allen[3], Peter J. Weddle,[2] Andrew M. Colclasure[2], Andrew N. Jansen[1]

[1]Chemical Sciences and Engineering Division, Argonne National Laboratory, Lemont, IL, USA

[2]Energy Conversion and Storage Systems Center, National Renewable Energy Laboratory, Golden, Colorado 80401, United States

[3]Computational Science Center, National Renewable Energy Laboratory, Golden, CO, USA

[*] Corresponding author

**Contact:** marco@anl.gov

**Abstract**

It is generally believed that silicon-based anodes for Li-ion batteries would benefit from stronger binders, as cyclic volume changes would not disrupt the cohesion of the composite electrode. Here, we put this belief to the proof by testing electrodes containing $SiO_x$ particles and an aromatic polyimide binder. We observe that the electrodes can stretch laterally by as much as 6% during the first cycle, indicating that internal stresses are high enough to induce plastic deformation on the copper current collector. Remarkably, no coating delamination is observed. Additional consequences were size-dependent: while pouch-cell-sized electrodes developed wrinkles, coin-cell-sized ones remained mostly smooth. We demonstrate that wrinkling of the current collector damages the electrode coating, inactivating $SiO_x$ domains and accelerating capacity fade. This size-dependent performance decay indicates that, in extreme cases, testing outcomes are highly dependent on scale. Novel battery materials may require testing at larger cell formats for complete validation.

## Introduction

Silicon-based materials and composites present high capacity for $Li^+$ storage, and their addition to the anodes of Li-ion batteries can increase cell energy.[1] The classical challenge with these materials is that they undergo volume changes upon cycling that can reach 300% on the particle level. Such dramatic structural breathing of a brittle material can initiate cracks that will deteriorate cell performance.[2] Changes in the architecture of Si-based materials, such as nanostructuring or encapsulation by hollow shells, have found success in mitigating the consequences of this cyclic dilation,[2-3] not by impeding the expansion but by accommodating it within the electrode structure.

The choice of binder is particularly important for Si electrodes. Carboxylated polymers are a common choice, as these moieties can bind covalently to the oxygenated groups at the silicon surface and form strong interfaces.[4] Nonetheless, many of these polymers lack the intrinsic mechanical strength needed to cope with the severe dimensional changes of Si. Aromatic polyimides have been investigated as an alternative, as they possess superb adhesive properties and a high density of polar groups to interact with the active material,[5] and many reports have shown how these polymers can improve the cycle life of Si-based materials.[3, 5] Nevertheless, too strong of an adhesion can have unintended consequences. Here we show that, as silicon expands during initial cycling, a strong polyimide binder will cause the electrode to stretch and wrinkle rather than delaminate. Although electrode integrity is successfully maintained, this mechanical deformation introduces additional modes of capacity decay. Furthermore, as we will show, these effects are heavily size-dependent, being nearly absent from smaller coin cells but present in larger prototypes, showing that the initial screening of new formulations could conceal critical problems.

In these studies, we used $SiO_x$ as the active material. Electrodes contained at least 60wt% of $SiO_x$, and 20wt% of an aromatic polyimide binder (see *Supporting Information* for details). Testing of this active material involves the formation of lithium silicates that are not fully reversible,[6] resulting in large initial coulombic inefficiency. This is commonly compensated for by introducing excess $Li^+$ into the cell through prelithiation.[7] Cells were prelithiated electrochemically, by first pairing the anode with Li metal at the appropriate cell format (coin or pouch cell) and running three full cycles (50-600 mV vs. Li). Prelithiated anodes were then harvested from the half-cells and paired with the desired cathodes to create full-cells.

Figure 1a-b shows the back and front of a pouch cell electrode (14.9 $cm^2$) containing 60wt% of $SiO_x$ and 17wt% of graphite (4.5 $mAh/cm^2$, 50-600 mV) after the prelithiation cycles. Severe wrinkling is visible in matching patterns on both the electrode coating and current collector, suggesting that the entire piece deformed as a whole. Quite remarkably, the current collector also underwent plastic deformation and stretched visibly by ~6% (Figure 1c). The stretching was non-uniform, being more pronounced in the regions where rippling was more prominent, inducing a slight curvature at the sides of the electrode (Figure 1d). It was also anisotropic, being somewhat smaller along the axis defined by the tab. The yield strength of the 10-μm-thick copper current collector should be higher than ~120 MPa,[8] indicating that large stresses must develop within the electrode to generate the observed plastic deformation. This elongation of the electrode likely stems from the combination of expanding $SiO_x$ particles and a highly adherent binder. Changes in particle volume can lead to debonding at the silicon-binder interface, which is commonly reported as a failure mode of Si-containing cells.[1-2, 9] In principle, a sufficiently strong binder could remain attached to the active material particles, transferring the resulting load to the current collector. The absence of delamination in the electrodes in Figure 1 suggests that the binding strength of the binder to both $SiO_x$ and Cu is larger than the yield strength of the current collector.

Despite the rippling of the coating, the electrode remained functional. Figure 1e shows cycling data for pouch cells containing the prelithiated anode and a NMC811 cathode, at two different ratios of areal capacity between the electrodes (N/P ratio). The cell with high N/P ratio exhibited no capacity fade after 400 cycles, while a higher anode utilization resulted in a cycle life of ~400.

A surprising observation was that some features of the mechanical behavior of the electrodes could not be observed in coin cells. When smaller electrodes (1.54 $cm^2$) were exposed to the same prelithiation protocols, a similar level of stretching could still be observed (Figure 1f) but with mild dimpling being observed instead of severe wrinkling (Figure 1g). Tests carried out using rectangular electrodes yielded similar observations (not shown). To evaluate whether the asymmetry induced by a tab could change the mechanical behavior, rectangular electrodes with a Cu tail were also tested (Figure 1h); an additional separator was placed behind the electrode and the tab was folded around it to serve as the sole point of electronic contact with the cell case. From these experiments, it appears that the occurrence of plastic deformation of the current collector is a property of the electrode, whereas other effects (such as rippling) also depend on sample dimension (Figure 1i).

To further investigate this phenomenon, we used Kirchhoff's thin plate theory (refs. [10-12]) to evaluate the analytical expression between buckling stress and cell size. We assumed that the bi-layer composite plate of silicon and copper is subjected to only uniaxial compression with no vertical deflection and moments on all domain boundaries, as shown in Figure 2a. This compressive stress arises during delithiation, as the silicon particles contract. The governing equations required to evaluate the buckling stress, along with the analytical solution, are given in Figure 2b. As the thickness of the copper foil is significantly smaller than the other two dimensions, we have incorporated the effect of length scale to determine the deformation behavior accurately.[13] The length scale here can be correlated with either the copper thickness or the grain size along the thickness direction. A detailed description of the governing equations, boundary conditions, and derivation to the analytical solution are provided as Supporting Information. Figure 2c illustrates the variation of buckling stress (i.e., the stress required to buckle/wrinkle) as a function of electrode dimension, exhibiting maximum values for smaller electrodes and a smooth decay as the initial size of the electrode increases. If we assume that copper is a perfectly plastic material, the stress within the current collector during cycling is estimated to fall over the dotted horizontal line shown in Figure 2c. Interestingly, below a critical electrode size ($L_c$), the buckling stress required to ripple the current collector is never achieved. In other words, rippling of the coating and current collector will not occur for electrodes that are smaller than $L_c$. Above this critical length, however, the buckling stress is considerably lower than the stresses experienced by copper foil (horizontal dotted line), causing larger electrodes to deform. This description agrees well with the experimental observations shown in Figure 1, in which a same level of plastic deformation of copper will result in a mechanical behavior that is more damaging in larger pouch cell electrodes.

The mechanical behavior also depended on the electrode loading. Figure 3a-b shows photographs of the back of two electrodes after testing, containing 70 wt% $SiO_x$ (no graphite) and with loadings of 1.3 and 2.2 mAh/cm$^2$ (50-600 mV). While the thinner electrode looked pristine (apart from marks from handling), the thicker one displayed irregular maze-like deformation patterns. A higher loading means more expanding particles per unit area, resulting in a larger total stress that is transferred by the binder to the current collector. Full-cell containing the prelithiated electrodes showed very distinct performance, with cycle lives of ~500 and ~900 for high and low loading, respectively (Figure 3c). Although other factors can contribute to this loading-dependent performance, it is possible that rippling can introduce defects to the electrode coating, facilitating cell decay.

Inspired by the observations above, we carried out tests using the high-loading $SiO_x$ electrode under identical conditions, but at two formats: coin cell (1.77 $cm^2$ anode) and pouch cell (14.9 $cm^2$). Since smaller electrodes exhibit no rippling (Figure 1f-i), this experiment can help determine whether deformation of the coating affects testing outcomes. Pouch cells were also tested under a stack pressure of ~30 psi, to replicate what the electrodes would experience in a typical coin cell.[14] The testing protocol involved 1C aging cycles, with RPTs taken every 100 cycles (including one full cycle at C/10 and another where pulses are applied to track impedance). Data for full-cells at both formats containing NMC811 and prelithiated anodes are shown in Figure 4, and the differences are striking. While the cells maintained similar capacity retention during the initial ~200 cycles, their behavior diverged later on (Figure 4a). The coin cells exhibited a mild change in slope, while the pouch cells underwent a precipitous capacity decay. This changes in slope caused the cycle life to decrease from >900 (in coin cells) to ~500 (in pouch cells). The rapid decay in capacity for pouch cells we also accompanied by a decrease in the coulombic efficiency of the 1C aging cycles, which reached as low as ~99.8% after 500 cycles (Figure 4b). As we will discuss below, this drop in coulombic efficiency is indicative of other aging mechanisms. The impedance showed an interesting behavior: while it shows a more linear increase for the coin cells, it has a discontinuity in the pouch cells (Figure 4c). This change in behavior coincides with the onset of the sudden hike in the rate of capacity fade, after ~200 cycles. The rapid capacity fade, impedance rise and low coulombic efficiency after cycle ~200 could be indicative of the occurrence of Li plating. Upon cell disassembly, visual confirmation was obtained that Li has plated on the pouch cell but not on the coin cell (Figure 4d-e).

Although Li plating in Si could be less likely due to anode potentials being higher than graphite, it has been reported to initiate due to capacity fade, which decreases the N/P ratio of the cell.[1, 15] The occurrence of Li plating solely in the cell format that led to rippling of the electrode coating (pouch cell) suggests that this disruption of the electrode coating could permanently decrease the accessible capacity. To quantify these losses, we built half-cells with electrodes harvested from coin cells and punched off the cycled pouch cell electrodes. By comparing their C/100 capacity with that of an unaged electrode, permanent capacity loss can be assessed. Figure 4f shows that minor losses of Si capacity were observed in coin cells (10%), while major losses were found in electrodes from pouch cells (30-70%). Furthermore, the capacity loss of electrodes harvested from pouch cells varied widely depending on the sampled region. We hypothesize that regions that experienced more mechanical disturbance are the ones with higher losses. Since wrinkling is non-uniform across the electrode surface, this could explain the variance in fade. Thus, there could be a direct link between mechanical deformation and loss of active material capacity. Additionally, close inspection of Figure 4e suggests that Li plating appears to have preferentially happened along the elevated ridges that form when the electrode wrinkles. Hence, the mechanical damage also appears to be contributing to Li plating through geometric heterogeneity.[16] Scheme 1 depicts how the rippling of the electrodes relate to accelerated aging. Losses of Si capacity due to coating damage can expose the anode to lower potentials, which can facilitate Li plating at moderate rates. To make matters worse, this overutilization of Si domains could induce further capacity losses, which increasingly favors Li plating. Additionally, the ripples over the electrode surface can create hotspots of reactivity, causing the electrode to experience Li plating conditions locally even when Li deposition is not to be expected.

In summary, binders with an unusually strong affinity to both active material and current collector can create unexpected issues in electrodes that experience dimensional changes during cycling. The adhesion of a polyimide binder avoided delamination of electrodes rich in $SiO_x$ during electrode expansion but also transferred the load to the current collector, resulting in plastic deformation (permanent stretching and rippling) of both electrode coating and current collector. This effect was more pronounced at higher electrode loadings and, crucially, larger electrode dimensions. The rippling of the electrode coating facilitated the occurrence of Li plating, both by inactivating the accessible Si capacity and by introducing heterogeneities across the electrode surface, and was observed in pouch cells but not in coin cells. As a result, the former vastly underperformed the smaller cell format. An important consequence of this effect is that, in cases as extreme as these, experimental outcomes will depend heavily on the size of cells under testing: certain aging modes may only be measurable when testing is performed with prototypes of the appropriate scale. This realization is critical when considering the initial validation of new battery technologies.

### *Acknowledgements*

This research was supported by the U.S. Department of Energy's Vehicle Technologies Office under the Silicon Consortium Project, directed by Brian Cunningham, and managed by Anthony Burrell. The submitted manuscript has been created by UChicago Argonne, LLC, Operator of Argonne National Laboratory ("Argonne"). Argonne, a U.S. Department of Energy Office of Science laboratory, is operated under Contract No. DE-AC02-06CH11357. This work was conducted in part by the Alliance for Sustainable Energy, LLC, the manager and operator of the National Renewable Energy Laboratory for the DOE under contract no. DE-AC36-08GO28308. The authors also acknowledge the efforts of Juliane Preimesberger and Jaclyn Coyle to provide the material properties for the analytical model. 

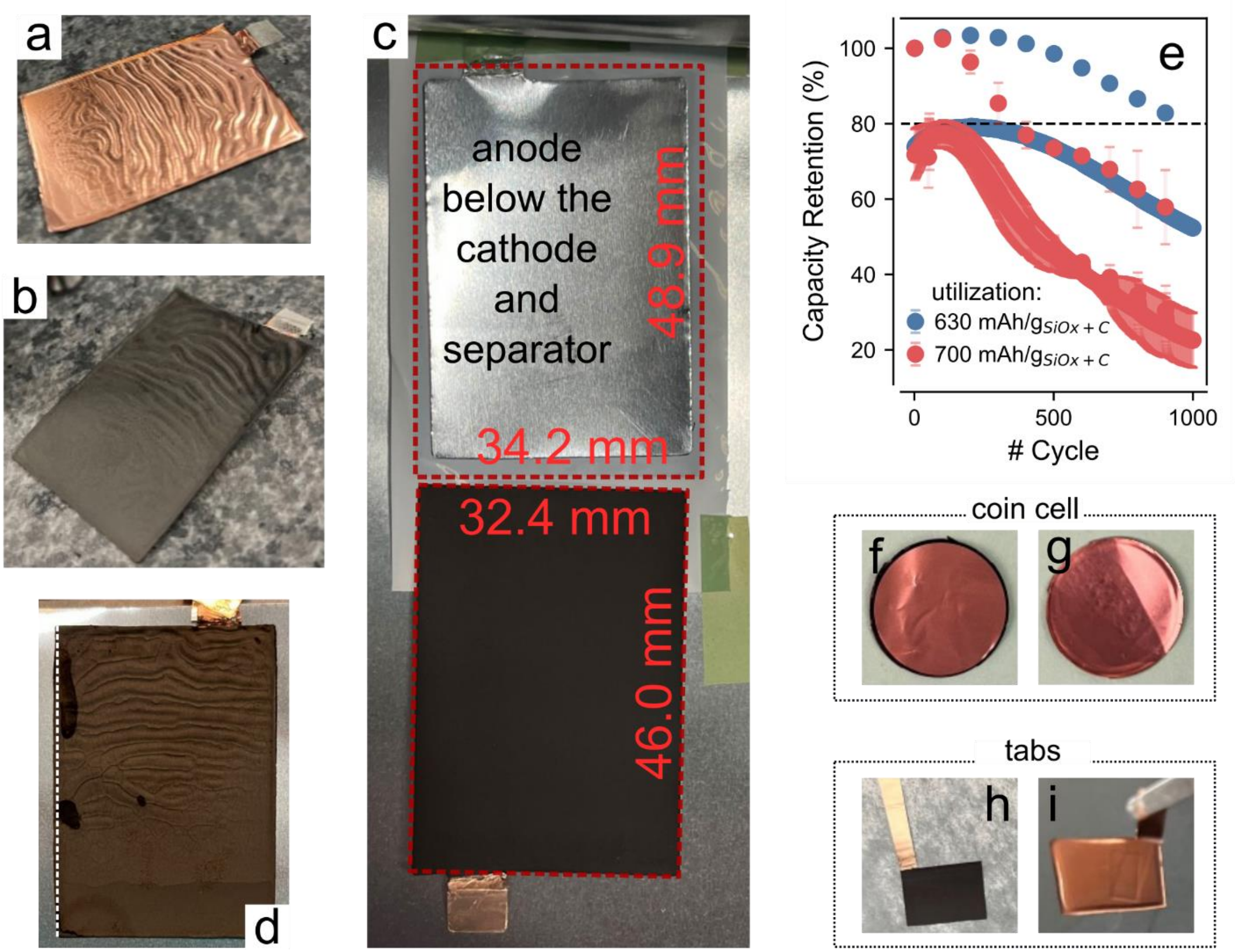

Figure 1. Pouch cells with 60 wt% $SiO_x$ and 17 wt% graphite. Back (a) and front (b) of an electrode after the prelithiation cycles ending with a delithiation to 600 mV vs. Li, showing the formation of ripples on both the Cu foil and the coating. c) Electrode stretching: a pristine electrode is shown upside down at the bottom, while the prelithiated electrode is on the top, underneath a cathode and a separator layer. Electrode edges are highlighted in red for reference. Dimensions before and after prelithiation indicate a ~6% anisotropic stretching. d) Same electrode as in panel b but with a white dashed line to highlight the curvature along the edges. e) Capacity retention of pouch cells using the stretched prelithiated electrodes, made with two different N/P ratios (indicated as anode capacity utilization). Initial full-cell areal capacities for high and low utilization were 4.2 and 3.7 mAh/cm$^2$. Error bars indicate two standard deviations and are too small to be visible for the low-utilization cells. Electrode stretching in coin cells: prelithiated 14-mm electrode under a 14-mm copper foil disc (f); back of the prelithiated electrode, showing much minor rippling than in pouch cells. Tab asymmetry in coin cells: pristine electrode with a tab (h) and back of the electrode after prelithiation (i); although the electrode stretched, the only marks on the copper are caused by compression of the tab, indicating that geometry is not a main driver for rippling.

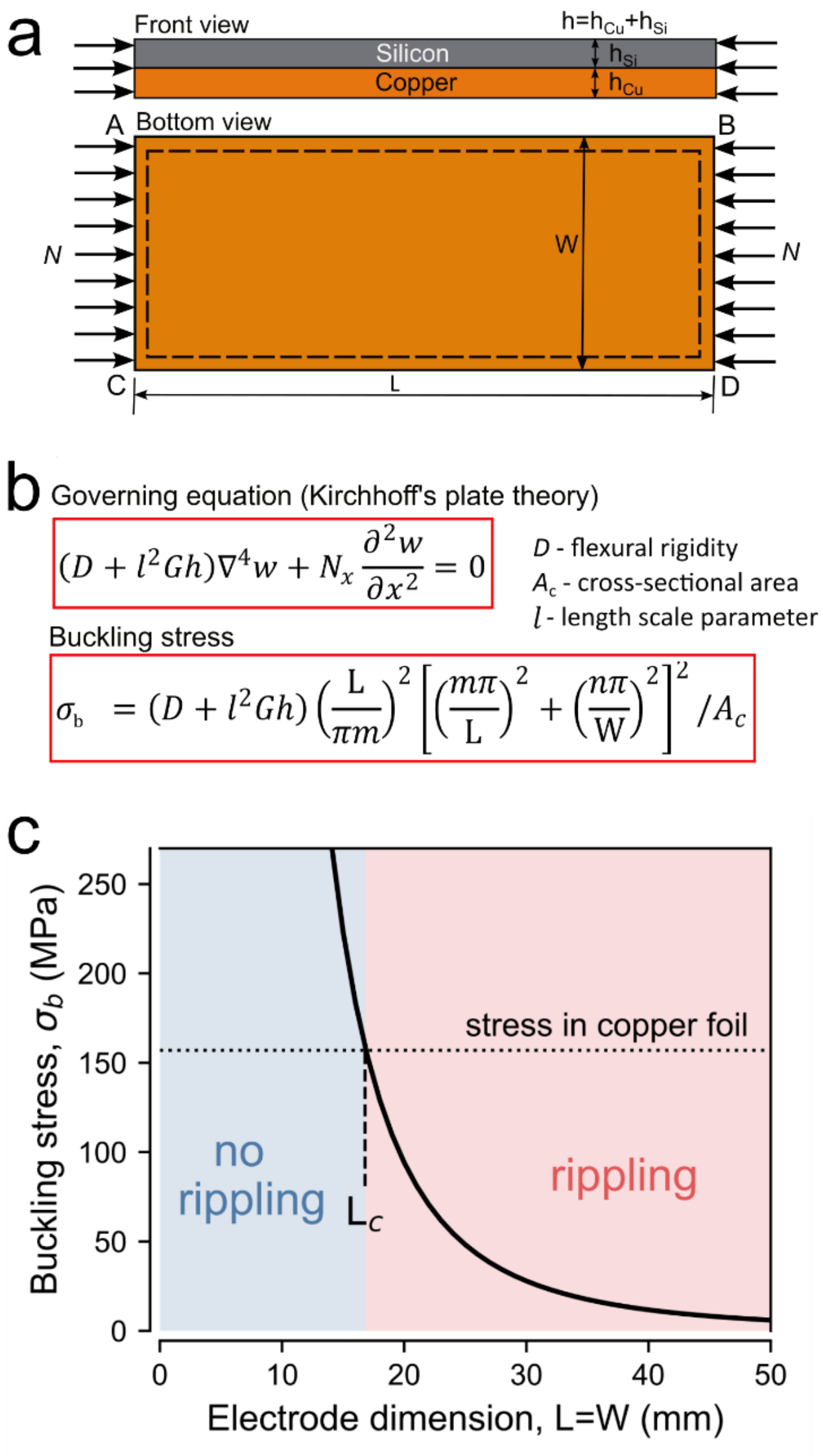

Figure 2. Modeling the plastic deformation of electrodes. a) Schematic illustration of the bilayer composite plate of silicon and copper subjected to uniaxial compression. b) Governing equation and its analytical solution for buckling/wrinkling stress. c) Variation of buckling stress (stress required to create ripples) as a function of the initial size of a square electrode. The estimated stress experienced by copper foil in the plastic regime (157 MPa) is marked by the dotted line. Electrodes smaller than the critical length $L_c$ will not exhibit ripples.

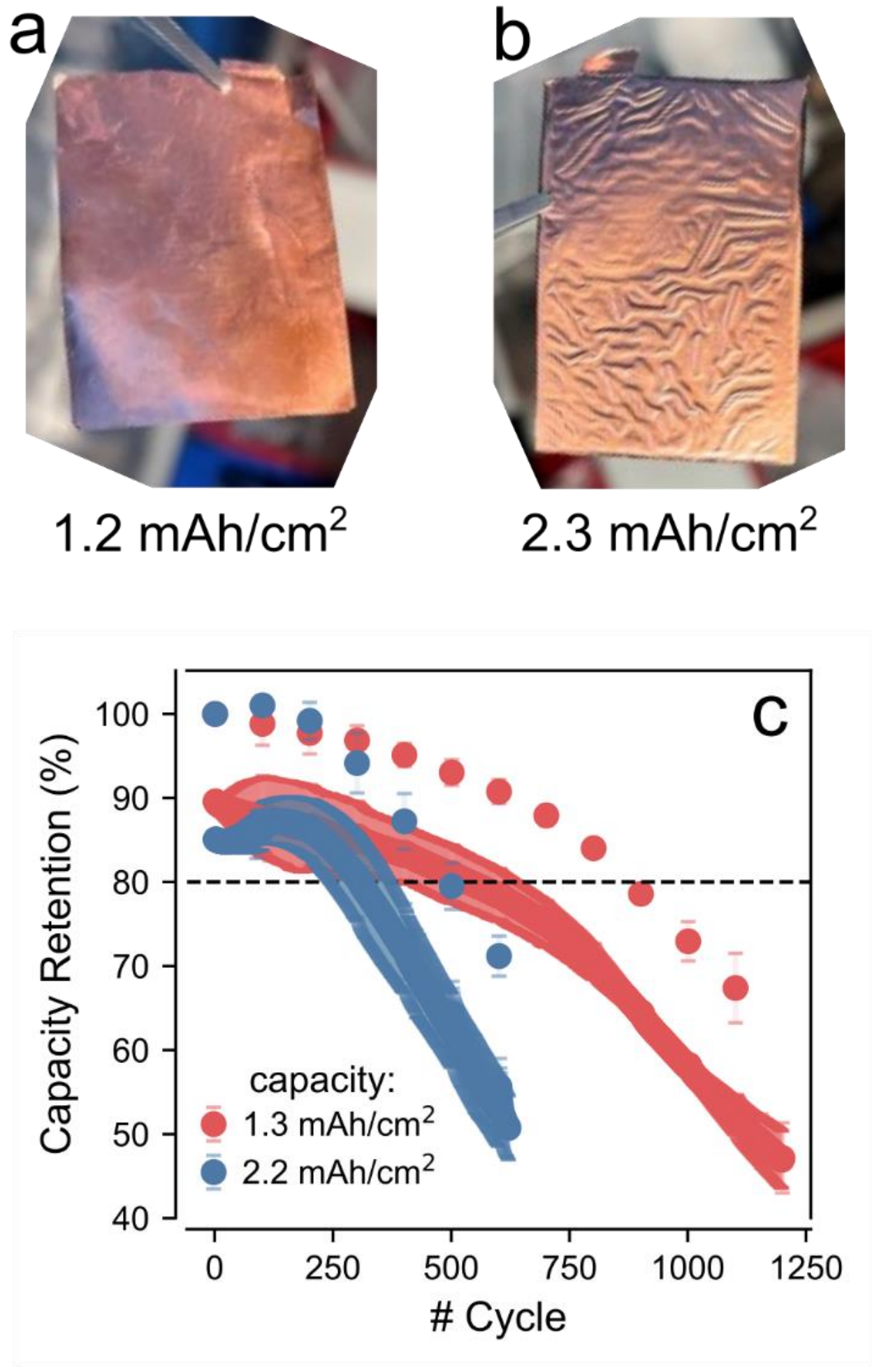

Figure 3. Pouch cells with 70 wt% $SiO_x$ with various areal loadings. Back of electrodes with lower (a) and higher loading (b) after testing. Rippling only becomes visible after a certain critical loading is achieved. c) Cycling data of full-cells using anodes of various loadings. The areal capacities indicate the initial capacity of the full-cell. Different NMC811 cathode were used in each case to achieve similar utilization of the anode. Performance decays much faster at higher loadings. Error bars indicate two standard deviations.

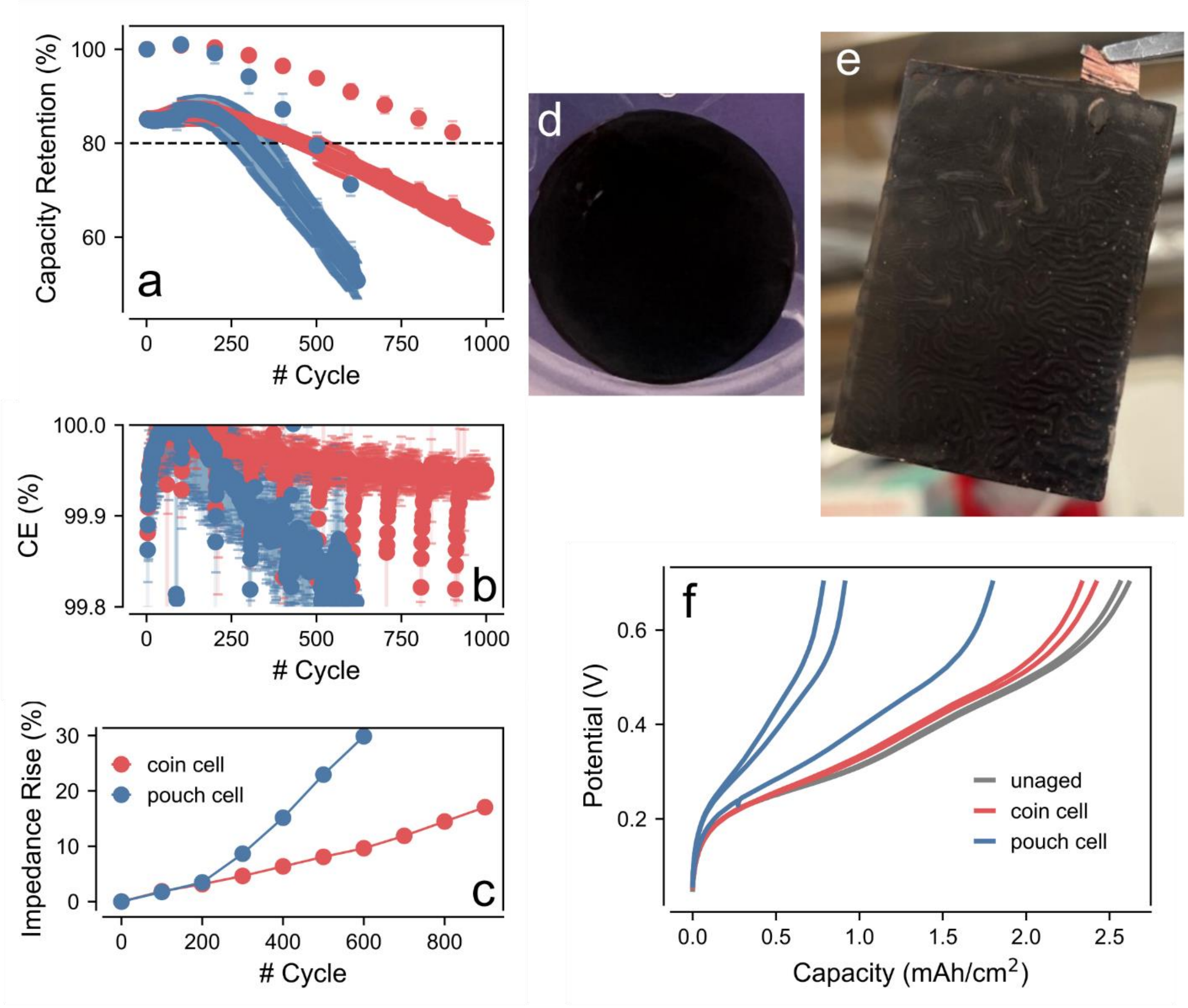

Figure 4. Testing electrodes with 70 wt% $SiO_x$ in both coin and pouch cells. Capacity retention (a) and coulombic efficiency (CE, panel b) of identical full-cells tested in either format. Error bars indicate two standard deviations. c) Impedance rise measured through high-rate pulses throughout the tests, showing rapid increase in pouch cells that coincides with the onset of faster decay in capacity and coulombic efficiency. The legend in panel c also applies to panels a and b. Photographs of electrodes at the end of testing on coin cells (d) and pouch cells (e). Li plating is only visible at the latter, and appears to follow the ridge-lines formed after rippling. f) Delithiation profiles collected at C/100 for half-cells containing unaged electrodes or electrodes harvested from the cycled coin and pouch cells. Minor permanent capacity fade is verified in coin cells, while major losses are found in pouch cells.

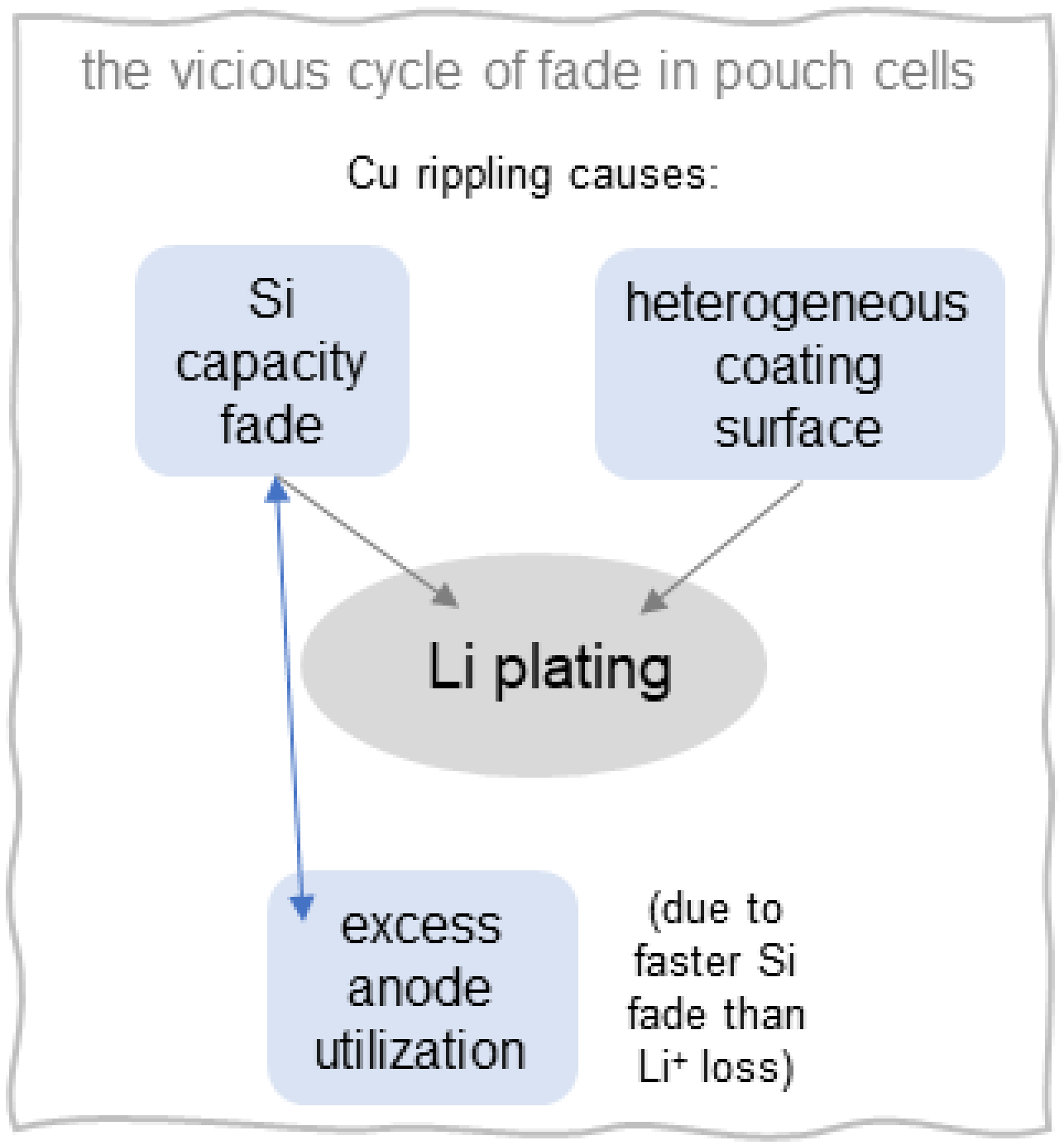

Scheme 1. Mechanical deformation and Li plating. Disruption of the electrode coating following plastic deformation of the Cu leads to permanent losses of Si capacity and creates a rough surface. Both these features can facilitate Li plating, by decreasing the effective anode potential during charging and by inducing heterogeneous reactivity, respectively. Furthermore, rapid deterioration of Si capacity could lead to higher levels of anode utilization (i.e., the anode is lithiated to lower potentials), which could further accelerate Si capacity loss.

*Supporting Information*

## Cell-format-dependent mechanical damage in silicon anodes

Marco-Tulio F. Rodrigues,[1*] Sathish Rajendran,[1] Stephen E. Trask,[1] Alison R. Dunlop,[1] Avtar Singh,[2] Jeffery M. Allen[3], Peter J. Weddle,[2] Andrew M. Colclasure[2], Andrew N. Jansen[1]

[1]Chemical Sciences and Engineering Division, Argonne National Laboratory, Lemont, IL, USA

[2]Energy Conversion and Storage Systems Center, National Renewable Energy Laboratory, Golden, Colorado 80401, United States

[3]Computational Science Center, National Renewable Energy Laboratory, Golden, CO, USA

[*] Corresponding author

**Contact:** marco@anl.gov

## Experimental details

The electrodes were fabricated at Argonne's Cell Analysis, Modeling and Prototyping (CAMP) Facility, and additional information can be found in Table S1. Pouch cell assembly was conducted in the xx3450 format, using a 14.1 $cm^2$ cathode and 14.9 $cm^2$ anode. For coin cells, 1.54 and 1.77 $cm^2$ discs were used, respectively. The separator was always Celgard 2500 and the electrolyte was 1.2M $LiPF_6$ in a 3:7 wt/wt mixture of ethylene carbonate (EC) and ethyl methyl carbonate (EMC), with 3 wt% fluoroethylene carbonate (FEC). All cells were "flooded" and contained an electrolyte volume equivalent to ~4x the total pore volume of electrodes and separator. All cells were assembled in a dry room (dew point < -50$^o$C). Electrochemical testing was performed at 30$^o$C.

Prelithiation was performed in the same cell format used for the final testing, by first constructing an anode vs. Li metal cell and performing three full cycles (50-600 mV at rates < C/10). After these initial cycles, cells were opened, lightly rinsed with dimethyl carbonate (DMC) and immediately used to assemble a full-cell with an appropriate cathode, and fresh separator and electrolyte. Initial cycles of full-cells were limited by the final delithiation capacity of the anodes during prelithiation, to avoid overcharging; final voltage cutoffs were defined based on the potentials reached by the cells during this initial charging. The final charge cutoff was 4.06 V for cells with 60% $SiO_x$ and 17% graphite, and 4.15 V for cells containing 70% $SiO_x$. Discharge cutoff was always 3 V. The 1C rates were defined as being 80% of the initial full-cell C/10 capacities, since this was the typical fraction of the nominal capacity that would be retained when cells were tested at 1C. Testing protocol involved reference performance tests (RPTs) spaced by 99 aging cycles at 1C. RPTs comprised a full cycle at C/10 and a cycle in which high-rate pulses were used to quantify cell impedance. In this latter cycles, cells were first fully charged, and then discharged at increments of 10% of the nominal cell capacity. After every partial discharge, cells rested for 1h, and then experienced a 3C discharge pulse for 10 seconds, a rest for 40 seconds and a 2.25C charge pulse for 10 seconds. A quantity resembling impedance can be calculated from the voltage change and the current applied during each pulse. Aging of pouch cells was always performed under a stack pressure of 26 psi to mimic what is experienced by electrodes in coin cells.[1] Prelithiation was carried out under 26 psi for pouch cells with 60% $SiO_x$ and 17% graphite, and under 4 psi for cells with 70 wt% $SiO_x$.

Cell teardown was performed in an argon-filled glovebox. Discs (1.54 $cm^2$) were punched out of the aged pouch-cell anodes and assembled into coin cells to assess electrode damage; for full-cells in a coin cell format, the original anode was carefully harvested from the cell. These harvested electrode cells also contained a Li metal counter electrode, and fresh electrolyte and separator, and were cycled 5 times between 50 and 700 mV at C/100. Cells with pristine electrodes were used as reference and were first cycled 5x at C/10 at this same voltage window.

Table S1. Electrodes used in this study. All electrodes were calendered to the indicated porosities. For the 70 wt% $SiO_x$ electrode, information for an additional electrode loading is provided in brackets.

| | |
|---|---|
| **60% SiOx, 17% graphite**<br>60 wt% Osaka $SiO_x$<br>3 wt% Timcal C45 carbon<br>17 wt% Superior Graphite 1506T<br>20 wt% Polyimide binder<br>Coating Thickness: 58 µm<br>Porosity: 41.6 %<br>Coating Loading: 7.13 mg/cm$^2$ | **70% $SiO_x$**<br>70 wt% Osaka $SiO_x$<br>10 wt% Timcal C45 carbon<br>20 wt% Polyimide binder<br>Coating Thickness: 17 µm (26 µm)<br>Porosity: 51.6 % (49.9 %)<br>Coating Loading: 2.74 mg/cm$^2$ (1.58 mg/cm$^2$) |
| **NMC811**<br>90 wt% Targray NMC811<br>5 wt% Timcal C-45<br>5 wt% Solvay 5130 PVDF Binder<br>Coating Thickness: 59 µm<br>Porosity: 34.5 %<br>Coating Loading: 15.81 mg/cm$^2$ | **NMC811**<br>96 wt% Targray NMC811<br>2 wt% Timcal C-45<br>2 wt% Solvay 5130 PVDF Binder<br>Coating Thickness: 72 µm<br>Porosity: 34.7 %<br>Coating Loading: 21.01 mg/cm$^2$ |

## Governing equations

A brief description of governing equations for the stability analysis of a thin bilayer rectangular plate composed of Si and Cu subjected to uniaxial compressive load is provided. The modified couple stress theory (MCST) (refs. [2-5]) is considered for which the constitutive relations are described as

$$\sigma_{ij} = \lambda \varepsilon_{kk} \delta_{ij} + 2\mu \varepsilon_{ij}, \tag{1}$$

$$\varepsilon_{ij} = \frac{1}{2}\left(u_{i,j} + u_{j,i}\right), \tag{2}$$

$$\chi_{ij} = \frac{1}{2}\left(\theta_{i,j} + \theta_{j,i}\right), \tag{3}$$

$$\theta_i = \frac{1}{2} e_{ijk} u_{k,j}, \tag{4}$$

$$m_{ij} = 2l^2 \mu \chi_{ij}, \tag{5}$$

$$\mu = G = \frac{E}{2(1+\nu)}, \tag{6}$$

$$\lambda = \frac{E\nu}{(1+\nu)(1-2\nu)}, \tag{7}$$

where $\sigma_{ij}$, $\varepsilon_{ij}$, $\varepsilon_{kk}$, $\chi_{ij}$, and $m_{ij}$ are the stress tensor, strain tensor, dilatational strain, symmetric rotation gradient tensor, and couple stress tensor, respectively; $\mu$ and $\lambda$ are Lamè constants; $l$ is the intrinsic length scale parameter of the material; $\theta_i$ is the rotation vector; $e_{ijk}$ is the permutation symbol; $E$ is the elastic modulus; and $\nu$ is Poisson's ratio, respectively.

Upon considering a thin plate of length $L$, width $W$ and thickness $h = h_{\mathrm{Si}} + h_{\mathrm{Cu}}$, the displacement field based on Kirchhoff thin plate theory (refs. [3-5]) is expressed as,

$$u(x, y, z) = -z\frac{\partial w}{\partial x}, \tag{8}$$

$$v(x, y, z) = -z\frac{\partial w}{\partial y}, \tag{9}$$

$$w(x, y, z) = w(x, y, t). \tag{10}$$

By utilizing Eq (1)-(10), the bending moment and shear force can be described as,

$$M_x = \int_h \sigma_{xx} z\, dz + \int_h m_{xy} z\, dz = \frac{-Eh^3}{12(1-\nu^2)}\left(\frac{\partial^2 w}{\partial x^2} + \nu\frac{\partial^2 w}{\partial y^2}\right) - l^2 Gh\left(\frac{\partial^2 w}{\partial x^2} - \frac{\partial^2 w}{\partial y^2}\right), \tag{11}$$

$$M_y = \int_h \sigma_{xx} z\, dz - \int_h m_{yx} z\, dz = \frac{-Eh^3}{12(1-\nu^2)}\left(\frac{\partial^2 w}{\partial y^2} + \nu\frac{\partial^2 w}{\partial x^2}\right) - l^2 Gh\left(\frac{\partial^2 w}{\partial y^2} - \frac{\partial^2 w}{\partial x^2}\right), \tag{12}$$

$$M_{xy} = M_{yx} = \int_h \sigma_{xy} z\, dz - \int_h m_{xx} z\, dz = \int_h \sigma_{yx} z\, dz + \int_h m_{yy} z\, dz = \frac{-Eh^3}{12(1-\nu^2)}\frac{\partial^2 w}{\partial x\, \partial y} - 2l^2 Gh\frac{\partial^2 w}{\partial x\, \partial y}, \tag{13}$$

$$Q_x = \int_h \sigma_{xz}\, dz, \tag{14}$$

$$Q_y = \int_h \sigma_{yz}\, dz, \tag{15}$$

The equilibrium equations can be expressed as follows:

$$\frac{\partial M_x}{\partial x} + \frac{\partial M_{yx}}{\partial y} = Q_x, \tag{16}$$

$$\frac{\partial M_y}{\partial y} + \frac{\partial M_{xy}}{\partial x} = Q_y, \tag{17}$$

$$\frac{\partial Q_x}{\partial x} + \frac{\partial Q_y}{\partial y} + q = 0. \tag{18}$$

Using Eqs. (16)-(18), the equilibrium equation for buckling/wrinkling of thin plate subjected to in-plane compressive load $N$ in $x$-axis direction is defined as,

$$(D + l^2 Gh)\nabla^4 w + N_x \frac{\partial^2 w}{\partial x^2} = 0, \tag{19}$$

in which the flexural rigidity $D$ is defined as

$$D = \frac{Eh^3}{12(1-\nu^2)}. \tag{20}$$

We have assumed that the deflection and moments along all edges of the thin plate is zero such that,

$$w = 0 \text{ and } M_n = 0 \text{ on ABCD} \tag{21}$$

(See Figure 2 of the main manuscript) The solution for a vertical deflection of the plate is sought as a product of two harmonic functions:

$$w(x, y) = w_0 sin \frac{m\pi x}{L} sin \frac{n\pi y}{W}, \tag{22}$$

where $m$ is the number of half-sine waves in the $x$-direction, $n$ is the number of half-sine waves in the $y$-direction, and $w_0$ is an amplitude of the deflection. It should be noted that the above equation can satisfy all corresponding boundary conditions. Using Eq. (22), the equilibrium equation for buckling/wrinkling (Eq. (19)) can be rewritten as,

$$\left\{(D + l^2 Gh)\left[\left(\frac{m\pi}{L}\right)^4 + 2\left(\frac{m\pi}{L}\right)^2\left(\frac{n\pi}{W}\right)^2 + \left(\frac{n\pi}{W}\right)^4\right] - N_x\left(\frac{m\pi}{L}\right)^2\right\} w_0 sin \frac{m\pi x}{L} sin \frac{n\pi y}{W} = 0. \tag{23}$$

The above differential equation is satisfied for all values of $(x, y)$ if its co-efficient satisfy,

$$N_x = (D + l^2 Gh)\left(\frac{L}{m\pi}\right)^2\left[\left(\frac{m\pi}{L}\right)^2 + \left(\frac{n\pi}{W}\right)^2\right]^2 \tag{24}$$

**Parameters:**

The predictive effective modulus of Si-Cu thin bilayer composite plate is assumed to be given by ref. [6]:

$$E = \frac{1+p^2q^4+2pq(2q^2+3q+2)}{(1+q)^3(1+pq)}E_{\mathrm{Cu}}, \tag{25}$$

where $p$ is the modulus ratio ($E_{\mathrm{Si}}/E_{\mathrm{Cu}}$) and $q$ is the thickness ratio ($h_{\mathrm{Si}}/h_{\mathrm{Cu}}$) of the two layers. The elastic modulus of Si and Cu are 19 GPa and 86 GPa, respectively. The Poisson's ratio of both Si and Cu are assumed to be 0.3. Yield strength of Cu is 157 MPa. The intrinsic length scale parameter $l = h_{\mathrm{Cu}}/2$. The solution parameters $m$ and $n$ are adopted as 5 and 1, respectively. The thickness of Si electrode and Cu foil are 14 $\mu m$ and 11 $\mu m$, respectively.

**Supporting References**